\documentclass[10pt, twocolumn, pre, aps, superscriptaddress, showpacs]{revtex4-1}
\usepackage{amsmath, graphicx, subfigure, tikz}
\begin{document}

\title{Spin-Glass Phases and Multichaos in the Ashkin-Teller Model}
\author{Alican Saray}
    \affiliation{Department of Physics, Bilkent University, Bilkent, Ankara 06800, Turkey}
    \affiliation{Department of Physics, Ohio State University, Columbus, Ohio 43210, USA}
\author{A. Nihat Berker}
    \affiliation{Faculty of Engineering and Natural Sciences, Kadir Has University, Cibali, Istanbul 34083, Turkey}
    \affiliation{Department of Physics, Massachusetts Institute of Technology, Cambridge, Massachusetts 02139, USA}

\begin{abstract}
The global phase diagram of the Ashkin-Teller spin glass is calculated in $d = 3$ spatial dimensions
by renormalization-group theory. Depending on the value of the positive or negative four-spin
interaction, qualitatively different topologies are found for the spin-glass phase diagram in the
usual variables of temperature and fraction of antiferromagnetic nearest-neighbor interactions. Two
different spin-glass phases occur. Both spin-glass phases are chaotic. One spin-glass exhibits phase
reentrance that is reverse from the reentrances seen in previous spin-glass phase diagrams.  Seven different phases: Ferromagnetic and antiferromagnetic, entropic ferromagnetic and entropic antiferromagnetic, spin-glass and entropic spin-glass, and disordered phases occur.  The entropic ferromagnetic phase unusually but understandably occurs at temperatures above one spin-glass phase. A random disorder line is identified and no phase transition occurs on this lie.  Our calculation is exact on the d = 3 hierarchical lattice and Migdal-Kadanoff approximate on the cubic lattice.
\end{abstract}
\maketitle

\section{Introduction: Chaotic Multi-Spin-Glass Order and Ashkin-Teller}

\subsection{Ashkin-Tellerized Spin Glass}

Spin-glass phases \cite{EdwardsAnderson,Toulouse,Sherrington,Paris1,Paris2,Paris3,Mydosh,Kawamura3}, resulting from mutually frustrating ferromagnetic versus antiferromagnetic (or left versus right chiral) interactions exhibit unsaturated order and signature chaotic rescaling behavior \cite{McKayChaos,McKayChaos2,McKayChaos4,BerkerMcKay}. These systems have been extensively studied for Ising spins.  On the other hand, the Ashkin-Teller model \cite{AT,Kadanoff0,Kecoglu}, a specifically doubled Ising model, has a variety of alternate local degrees of freedom, resulting, as seen below, in added richness to spin-glass phenomena. The spin-glass version of the Ashkin-Teller model is defined by the
Hamiltonian
\begin{equation}
- \beta {\cal H} = \sum_{\left<ij\right>} \, -\beta {\cal H}_{ij}= \sum_{\left<ij\right>} \, [J (s_i s_j +  t_i t_j) + M s_i s_j t_i t_j ]
\end{equation}
where $\beta = 1/k_B T$ is the inverse temperature, at each site $i$ there are two Ising spins $s_i = \pm 1, t_i = \pm 1$, and the sum is over all interacting quadruples of spins on nearest-neighbor pairs of sites. The two-spin interaction $J_{ij}$ is ferromagnetic $(J_{ij} = + J > 0)$ with probability $1-p$ and antiferromagnetic $(J_{ij} = -J < 0)$ with probability $p$. Under renormalization group, we follow the distribution of the transfer matrix $\textbf{T}_{ij}(s_i,t_i;s_j,t_j) = exp(-\beta{\cal H}_{ij}(s_i,t_i;s_j,t_j))$, with Eq.(1) being the initial condition of the renormalization-group flows. Thus, the four-spin interaction also acquires randomness (and chaos, as seen below). At each site $i$, another spin can be defined, $\sigma_i = s_i t_i$ , so that the product of any two of $(s_i,t_i,\sigma_i)$ gives the third one.  We shall see below that spin-glass quenched randomness in $J_{ij}$ induces spin-glass order in $(s_i,t_i)$, as expected, but also distinct spin-glass order in $\sigma_i$.

\begin{figure}[ht!]
\centering
\includegraphics[scale=0.40]{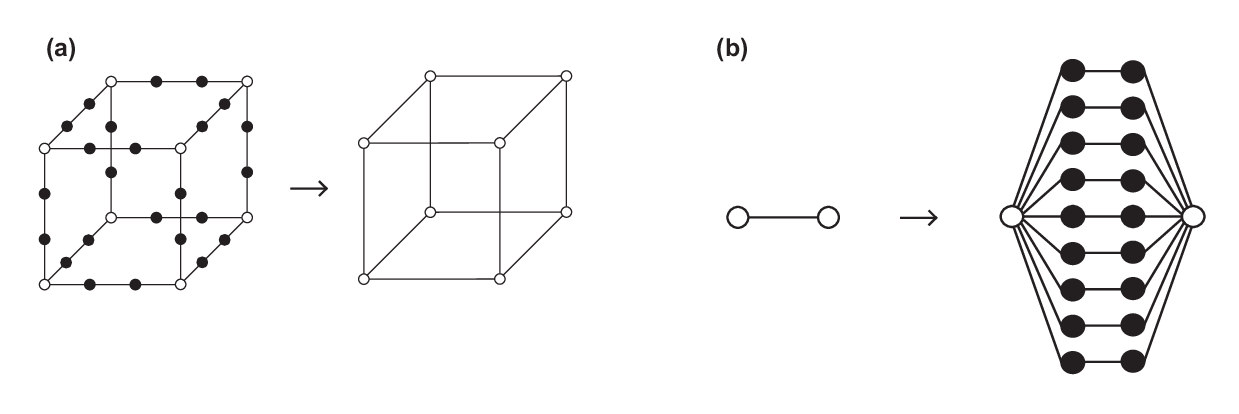}
\caption{Exactly solved hierarchical model and Migdal-Kadanoff: (a) The Migdal-Kadanoff approximate renormalization-group transformation on the $d=3$ cubic lattice. Bonds are removed from the cubic lattice to make the renormalization-group transformation doable.  The removed bonds are compensated by adding their effect to the decimated remaining bonds.  (b) A hierarchical model is constructed by self-imbedding a graph into each of its bonds, \textit{ad infinitum}.\cite{BerkerOstlund}  The exact renormalization-group solution proceeds in the reverse direction, by summing over the internal spins shown with the dark circles.  Here is the most used, so called "diamond" hierarchical lattice \cite{BerkerOstlund,Kaufman1,Kaufman2,BerkerMcKay}.  The length-rescaling factor $b$ is the number of bonds in the shortest path between the external spins shown with the open circles, $b=3$ in this case. The volume rescaling factor $b^d$ is the number of bonds replaced by a single bond, $b^d=27$ in this case, so that $d=3$.}
\end{figure}

\begin{figure*}[ht!]
\centering
\includegraphics[scale=0.51]{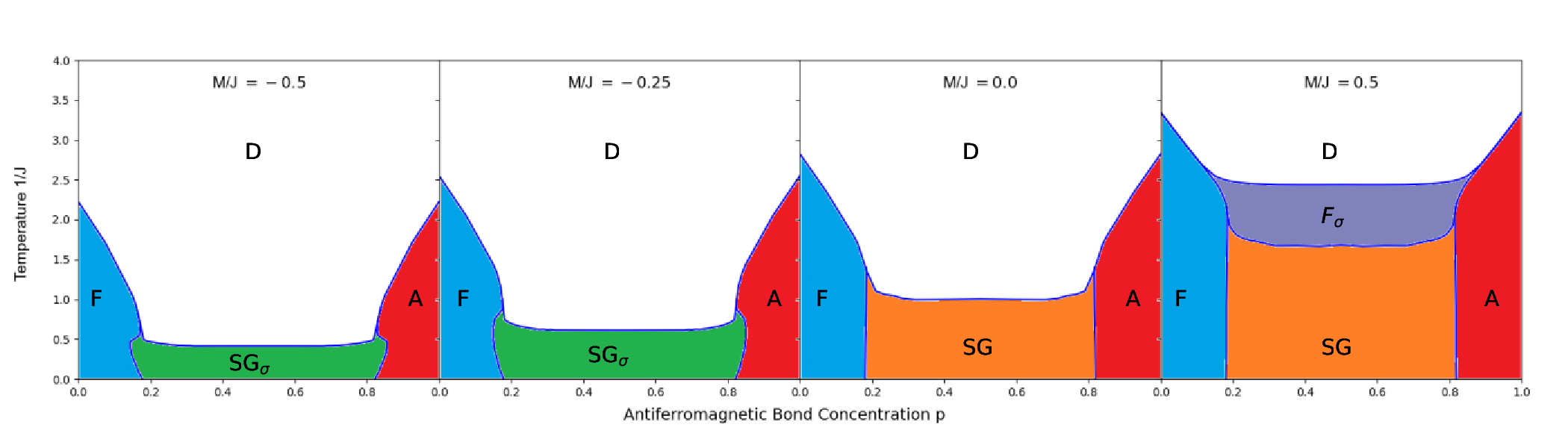}
\caption{Calculated cross-sections of the global phase diagram of the Ashkin-Teller spin glass, in the usual variables of temperature $1/J$ versus fraction $p$ of antiferromagnetic nearest-neighbor interactions, for different values of the 4-spin interaction $M/J$ . The division by $J$ is made to scale out the temperature in the dimensionless interactions, as seen in Eq.(1). In all cross-sections, the disordered phase $(D)$ occurs at high temperature. The ferromagnetic $(F)$ and antiferromagnetic $(A)$ phases, occurring at respectively high and low $p$, are respectively aligned and antialigned in the $s_i$ spins and, separately, $t_i$ spins. The composite ferromagnetic $(F_\sigma)$ and antiferromagnetic $(A_\sigma)$ phases are respectively aligned and antialigned in the $\sigma_i = s_i t_i$ spins. Since for each sign of $\sigma_i$, two combinations of $(s_i,t_i)$ are possible, these phases have a ground-state entropy \cite{BK1,BK2} of $ln 2$ per site. This entropy explains the unusual position of this ferromagnetic $F_\sigma$ phase: \textit{above}, in temperature, a spin-glass phase. The spin-glass phases $SG$ and $SG_\sigma$ are respectively spin-glass phases in the spins $(s_i,t_i)$ and $\sigma_i $, the latter again being more entropic by $ln 2$ per site. A phase reentrance is seen in the new spin-glass phase that is reverse from the reentrance seen \cite{ReentSG} in the usual spin-glass phase diagrams. As the disorder line $M/J = -1$ is approached from above, the phase transitions disappear, as for example seen in Fig. 3 and consistently with the disorder-line characteristic.  For $M/J<-1$, a phase transition occurs to the $A_\sigma$ phase, at the temperature $1/J$  shown in Fig. 3 as function of $M/J$ independent of $p$.  }
\end{figure*}

\begin{figure}[ht!]
\centering
\includegraphics[scale=0.72]{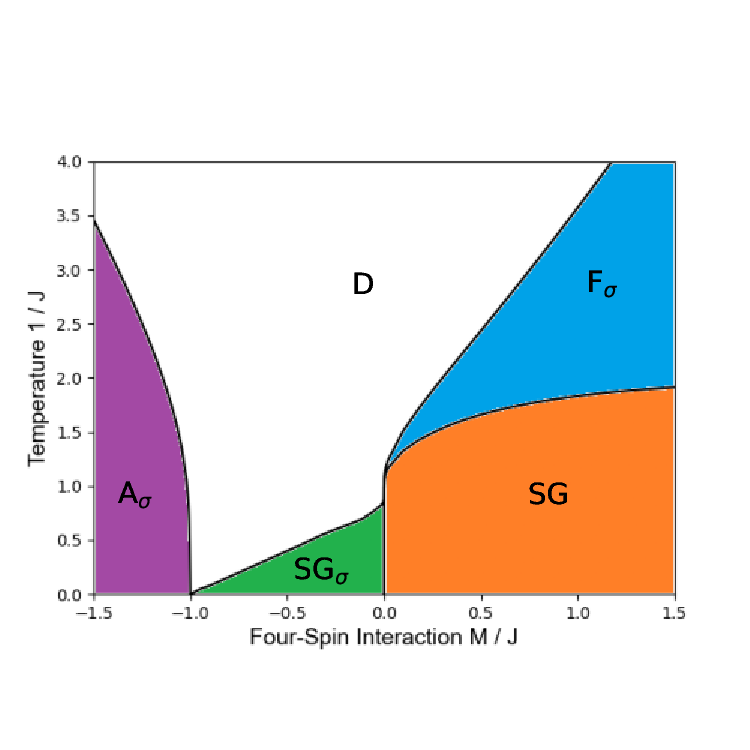}
\caption{The calculated phase transition temperatures $1/J$, at $p = 0.5$, as a function of the 4-spin interaction $M/J$ with temperature scaled out. The high-entropy phases, namely the composite ferromagnet, antiferromagnet and the two spin glasses, are seen. At $M/J = -1$, a random quasi disorder line is \textit{a priori }deduced (Sec. IB) and indeed, as seen in this figure, no phase transition occurs.}
\end{figure}

\subsection{Random Disorder Line}

Of the 16 possible states of the nearest-neighbor quadruple of $(s_i,t_i;s_j,t_j)$, at $M=-J$, for both $J_{ij}=\pm J$, 12 states have energy $\beta{\cal H}_{ij}(s_i,t_i;s_j,t_j)) = M$ and 4 states have energy $-3M$.  Thus, for $M<0$, the ground-state energy is highly degenerate \cite{BK1,BK2} and no ordering can be expected.  In the phase diagram, the line $M/J = -1$ is an \textit{a priori} quasi (because not all states are degenerate) disorder line, here occurring for a quenched random system.  Indeed, our calculations show that no ordering occurs at $M/J = -1$.

\section{Method: Exact Hierarchical Solutions}

\subsection{Exactly Solved d=3 Hierarchical Model}

Hierarchical models \cite{BerkerOstlund,Kaufman1,Kaufman2,BerkerMcKay} are exactly solvable microscopic models that are widely used.\cite{Sponge,CubicSG,Feigenbaum,Akimenko,Clark,Kotorowicz,ZhangPP,Jiang,Derevyagin2,Chio,Teplyaev,Myshlyavtsev,Derevyagin,Shrock,Monthus,Sariyer} The construction of a hierarchical model is illustrated in Fig. 1(b) [16]. The exact renormalization-group solution of the hierarchical model proceeds in the direction reverse to its construction (Fig. 1), by summing over the internal spins shown in the figure with the dark circles. In addition to phase transitions \cite{BerkerOstlund,Kaufman1,Kaufman2,BerkerMcKay} and spin-glass chaos \cite{McKayChaos,McKayChaos2,McKayChaos4,BerkerMcKay}, most recently hierarchical models have been developed for the classification and clustering of complex phenomena, including multicultural music and brain electroencephalogram signals \cite{MusicBrain,Pekcan}.

\subsection{Physical Justification of Migdal-Kadanoff}

The Migdal-Kadanoff approximation \cite{Migdal,Kadanoff} is a physically inspired, widely applicable and easy to implement, most used and most successful renormalization-group approximation. For example, the physical cubic lattice (Fig. 1a) cannot be directly renormalization-group transformed. Thus, some of the bonds are removed to make this transformation doable and the transformation is achieved by decimation, namely summing, in the partition function, over the intermediate spins (in black in Fig. 1a). However, the erstwhile removal of some bonds weakens the connectivity of the system. Thus, the effect of every bond removed is added to the decimated bonds. In addition to the application to physical systems such as surface systems \cite{BOP} and high-temperature superconductivity \cite{highTc}, the Migdal-Kadanoff approximation gives the correct lower-critical dimensions and ordering behaviors for all spin systems, including complex quenched-random systems.

As seen in Figs. 1, the Migdal-Kadanoff approximate solution is algebraically identical with the exact solution of a hierarchical model, which makes the Migdal-Kadanoff approximation a physically realizable approximation, as is used in turbulence \cite{Kraichnan}, polymers and gels \cite{Flory,Kaufman}, electronic systems \cite{Lloyd}, and therefore a robust approximation (e.g., it will never yield a negative entropy).

\subsection{Transfer Matrix Implementation}

The transformation is best implemented algebraically by writing the transfer matrix between two neighboring sites, namely,
\begin{multline}
\textbf{T}_{ij} \equiv e^{-\beta {\cal H}_{ij}} = e^{J (s_i s_j + t_i t_j) + M s_is_jt_it_j} = \\
\left(
\begin{array}{cccc}
e^{2J+M} & e^{-M} & e^{-M} & e^{-2J+M} \\
e^{-M} & e^{2J+M} & e^{-2J+M} & e^{-M} \\
e^{-M} & e^{-2J+M} & e^{2J+M} & e^{-M} \\
e^{-2J+M} & e^{-M} & e^{-M} & e^{2J+M}\end{array} \right),
\end{multline}
where the consecutive states are $(s_i,t_i) = (+1,+1),(+1,-1),(-1,+1),(-1,-1)$. The decimation step consists in matrix-multiplying $b$ transfer matrices. The bond-moving step consists in multiplying, element by element at each position in the matrix, the elements of $b^{d-1}$ transfer matrices. Here b = 3 is the length-rescaling factor of the renormalization-group transformation.  With no loss of generality, after each operation, the new transfer matrix is divided by its largest element. This amounts to subtracting a constant from the Hamiltonian and prevents computational blow-ups as the ordered phase sinks are approached. Under renormalization-group transformation, the inner and outer $2\times2$ structure of the transfer matrix is preserved, which means that the flows are in terms of $(J,M)$ and the symmetries of the Ashkin-Teller interaction are preserved.

\subsection{Spin-Glass Quenched Randomness}

As seen aboved, $b^d$ transfer matrices give one renormalized transfer matrix. For the renormalization-group transformation defined above, quenched randomness is taken into account as follows. A distribution of 10,000 transfer matrices is monitored, starting with the initial conditions specified after Eq.(1). From this distribution, $b^d$ randomly chosen transfer matrices are assembled, as described above, to give a renormalized transfer matrix.  For each renormalization-group step, This is repeated 10,000 times, to generate the renormalized distribution.

\begin{table*}

\begin{tabular}{c}
\multicolumn{1}{c}{Renormalization-Group Sinks of the Ashkin-Teller Spin Glass} \\
\end{tabular}

\begin{tabular}{c c c c c c c c c  c c}

\hline

\vline & $\begin{pmatrix} 1 & 0 & 0 & 0 \\ 0 & 1 & 0 & 0 \\ 0 & 0 & 1 & 0 \\ 0 & 0 & 0 & 1 \end{pmatrix}$  &\vline  & $\begin{pmatrix} 0 & 0 & 0 & 1\\ 0 & 0 & 1 & 0 \\ 0 & 1 & 0 & 0 \\ 1 & 0 & 0 & 0 \end{pmatrix}$  &\vline  &  $\begin{pmatrix} 1 & 0 & 0 & 1 \\ 0 & 1 & 1 & 0 \\ 0 & 1 & 1 & 0 \\ 1 & 0 & 0 & 1 \end{pmatrix}$  &\vline &$\begin{pmatrix} 0 & 1 & 1 & 0 \\ 1 & 0 & 0 & 1 \\ 1 & 0 & 0 & 1 \\ 0 & 1 & 1 & 0 \end{pmatrix}$ & \vline & $\begin{pmatrix} 1 & 1 & 1 & 1 \\ 1 & 1 & 1 & 1 \\ 1 & 1 & 1 & 1 \\ 1 & 1 & 1 & 1 \end{pmatrix}$ &\vline  \\
\hline
\vline & Ferromagnetic &\vline & Antiferromagnetic  &\vline & $\sigma$ Ferromagnetic &\vline & $\sigma$ Antiferromagnetic  &\vline & Disordered &\vline   \\
\hline

\end{tabular}

\caption{Under repeated renormalization-group transformations, the phase diagram points of the phases of the Ashkin-Teller spin glass flow to the sinks shown on this Table giving the exponentiated nearest-neighbor Hamiltonians. The (chaotic) sinks of the two spin-glass phases are given in Fig. 4.}
\end{table*}

\begin{figure}[ht!]
\centering
\includegraphics[scale=0.18]{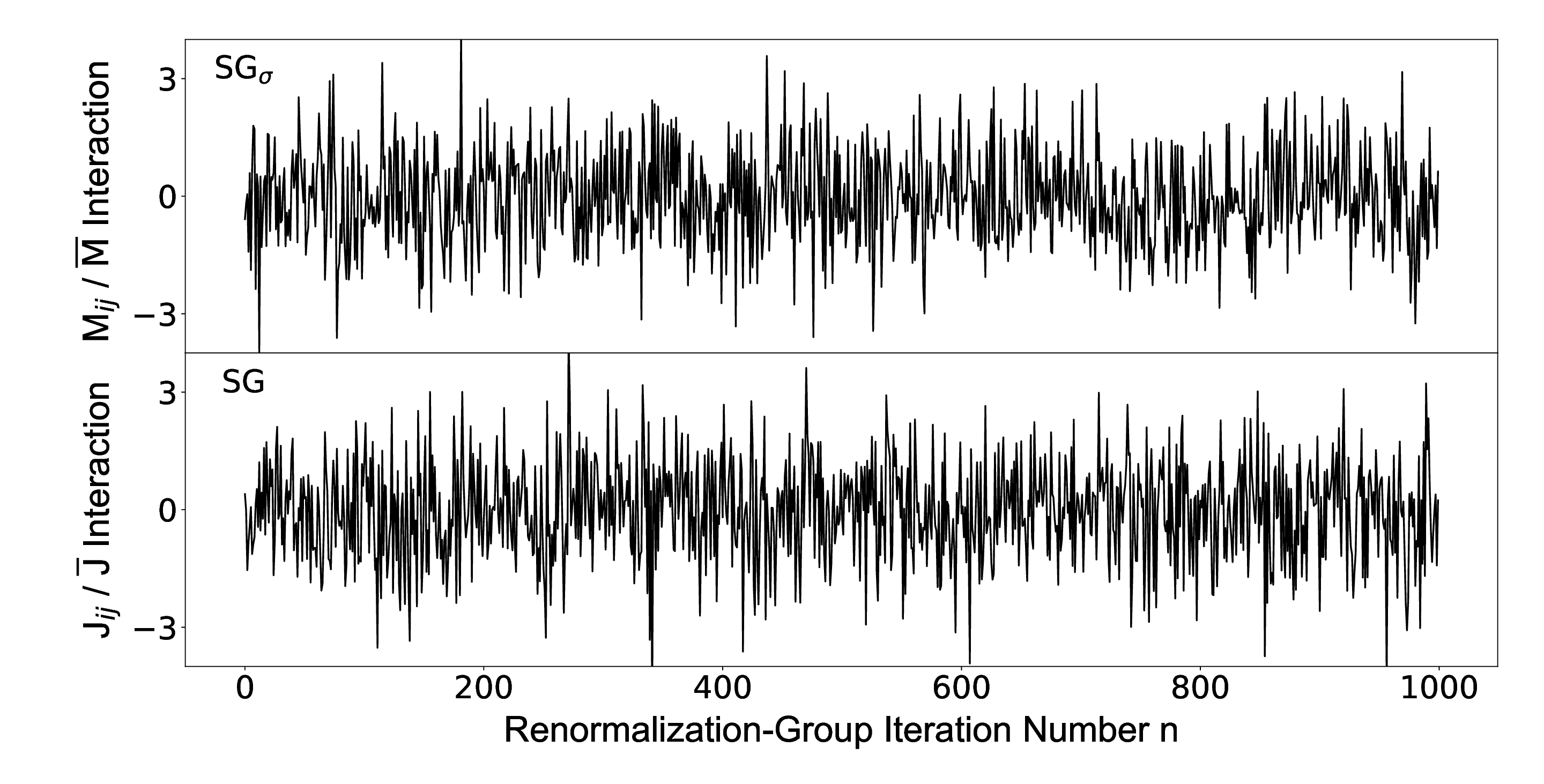}
\caption{Calculated chaotic renormalization-group trajectories of the two spin-glass phases.}
\end{figure}

\begin{figure}[ht!]
\centering
\includegraphics[scale=0.33]{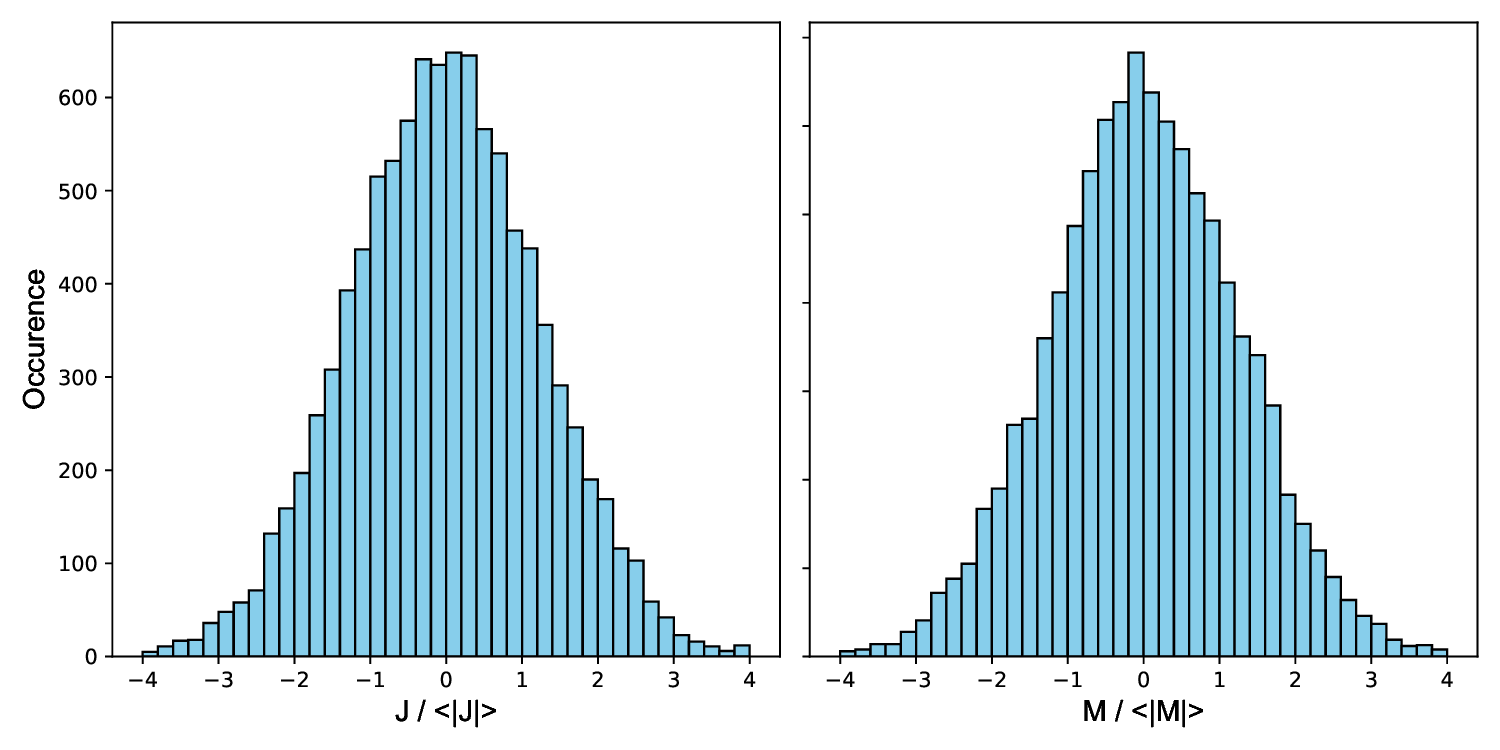}
\caption{Fixed distributions of the two spin-glass phases.}
\end{figure}

\begin{figure}[ht!]
\centering
\includegraphics[scale=0.42]{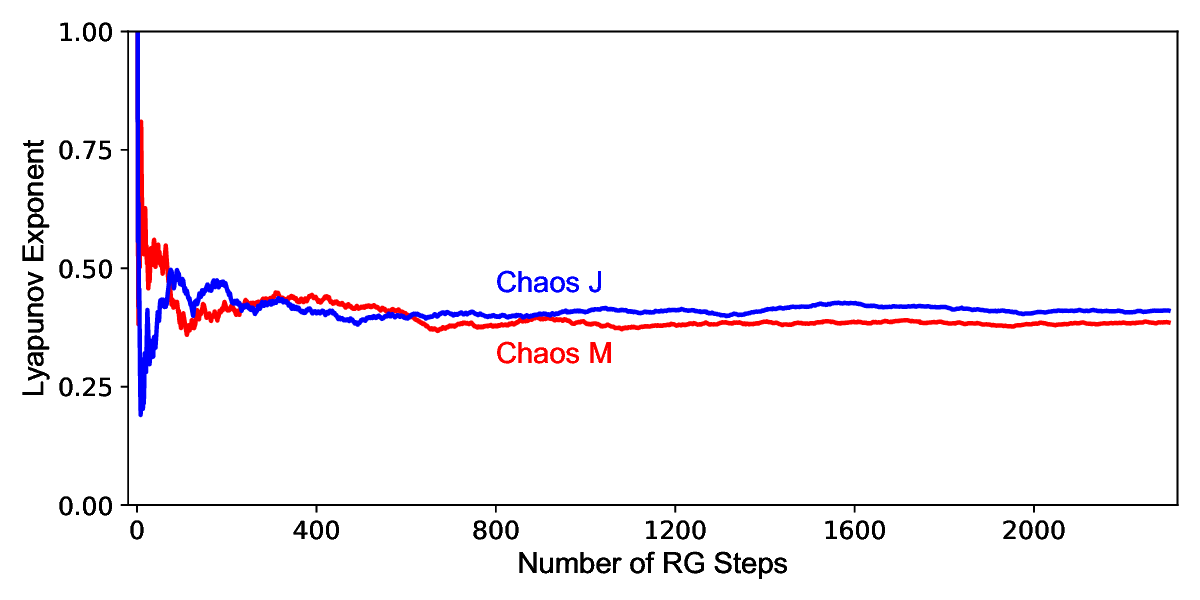}
\caption{Calculation, as in Eq.(3), of the Lyapunov exponents of $J$ chaos and $M$ chaos of the two spin-glass phases.}
\end{figure}

\section{Results: Global Phase Diagram, High-Temperature Entropic Ferromagnetism, Reverse Reentrance}

\subsection{Renormalization-Group Phase Sinks and Spin-Glass Fixed Distributions}

The global phase diagram of the Ashkin-Teller spin glass is given in Figs. 2 and 3. The points in each phase, under repeated renormalization-group transformations, flow to the totally stable fixed point, namely sink, of that phase.\cite{BerkerWor}  The values (or the distribution of values in the case of the spin-glass phases) of the transfer matrix at the sink epitomize the ordering of the phase that is the basin of attraction of the sink.  Thus, at the sink of the disordered phase $(D)$, all elements of the transfer matrix are unity.  At the sink of the ferromagnetic $(F)$ and antiferromagnetic $(A)$ phases, the diagonal and antidiagonal elements, respectively, are equal to unity, all other elements are zero.  The sink of the composite ferromagnetic phase $F_\sigma$ has the diagonal and anti-diagonal elements equal to unity and all other elements zero.  The sink of the composite antiferromagnetic phase $A_\sigma$ has the opposite of the $F_\sigma$ sink.  The sink of each spin-glass phases has a fixed distribution of positive and negative interactions, with the absolute values diverging.  In the case of the spin-glass phase $SG$, the 4-spin interaction $M$ renormalizes to zero as the fixed distribution is approached and the attractive fixed distribution and chaos are both in terms of the 2-spin interaction $J$.  Conversely, in the case of the spin-glass phase $SG_\sigma$, the 2-spin interaction $J$ renormalizes to zero as the fixed distribution is approached and the attractive fixed distribution and chaos are both in terms of the 4-spin interaction $M$.  The phase sinks and fixed distributions are respectively shown in Table I and Fig. 4.

\subsection{Global Phase Diagram and Entropic High-Temperature Ferromagnetism}

Calculated cross-sections of the global phase diagram of the Ashkin-Teller spin glass, in the usual variables of temperature $1/J$ versus fraction $p$ of antiferromagnetic nearest-neighbor interactions, for different values of the 4-spin interaction $M/J$, are shown in Fig. 2. The division by $J$ is made to scale out the temperature in the dimensionless interactions, as seen in Eq.(1). In all cross-sections, the disordered phase $(D)$ occurs at high temperature. The ferromagnetic $(F)$ and antiferromagnetic $(A)$ phases, occurring at respectively high and low $p$, are respectively aligned and antialigned in the $s_i$ spins and, separately, $t_i$ spins. The composite ferromagnetic $(F_\sigma)$ and antiferromagnetic $(A_\sigma)$ phases are, respectively, aligned and anti-aligned in the $\sigma_i = s_i t_i$ spins. Since for each sign of $\sigma_i$, two combinations of $(s_i,t_i)$ are possible, these phases have a ground-state entropy \cite{BK1,BK2} of $ln 2$ per site. This ground-state entropy explains the unusual position of this ferromagnetic phase: \textit{above}, in temperature, of a spin-glass phase, which are also known to unsaturate as zero temperature is reached and thus have ground-state entropy. \cite{BerkerYesil} Similarly, the composite spin-glass phase $(SG_\sigma)$ has an ground-state entropy of $ln 2$ per site, over the interacting spin-glass entropy.   As the disorder line $M/J = -1$ is approached from above, the phase transitions disappear, as for example seen in Fig. 3 and consistently with the disorder-line characteristic.  For $M/J<-1$, a phase transition occurs to the $A_\sigma$ phase, at the temperature $1/J$  shown in Fig. 3 as function of $M/J$ independent of $p$.

\subsection{Reverse Reentrance and Double Chaos}

The spin-glass phases $SG$ and $SG_\sigma$ are respectively spin-glass phases in the spins $(s_i,t_i)$ and $\sigma_i $. Thus, two different spin-glass phases are seen.  In the $SG_\sigma$ phase, there are again 2 combinations of $(s_i,t_i)$ for each $\sigma_i $ value, so that this phase has an extra entropy of $ln 2$ per site over the $SG$ phase.   Fig. 3 shows the continuous evolution of the entropic phases, namely the composite ferromagnetic $F_\sigma$ and antiferromgnetic $A_\sigma$ phases, and the two spin-glass phases $SG$ and $SG_\sigma$, as a function of $M/J$.

The phase diagram cross-sections in Fig. 2 show reentrance.  Reentrance is the disappearance and reappearance of a phase in the phase diagram, as temperature is lowered.\cite{Cladis,Netz,Garland,Walker,Caflisch,transverse,Ilker1,Mann1,Mann2}  Thus, for $M<0$, as temperature is lowered, the ferromagnetic phase yields to the spin-glass $SG_\sigma$ phase, and the same ferromagnetic phase reappears at lower temperature.  In fact, this is the reverse of the reentrance usually seen in spin-glasses \cite{ReentSG}, where, as temperature is lowered, the ferromagnetic phase appears and then disappears, as seen in Fig. 2 here for $M>0$.

The chaotic renormalization-group trajectories characterizing the spin-glass phases are shown in Fig. 4. The renormalization-group trajectories starting in the spin-glass phases do not go to a sink with a single transfer matrix as the ones for the other phases in Table I, but to a distribution of transfer matrices, shown in Fig. 5 for each spin-glass phase.  Two exponents characterize the chaotic sinks of the spin-glass phases.  The Lyapunov exponent \cite{Collet,Hilborn}
\begin{equation}
\lambda = \lim _{n\rightarrow\infty} \frac{1}{n} \sum_{k=0}^{n-1} \ln \Big|\frac {dx_{k+1}}{dx_k}\Big|\,,
\end{equation}
where $x_k = J_{ij}/\overline{J}$ and $x_k = M_{ij}/\overline{M}$ at step $k$ of the renormalization-group trajectory and $\overline{J}$ and $\overline{M}$ are the average of the absolute value in the quenched random distribution, respectively in the spin-glasses $SG$ and $SG_\sigma$. The location $ij$ and the renormalized locations overlaying it are included in the summation, which readily converges. Thus, we calculate the Lyapunov exponents by discarding the first 100 renormalization-group steps (to eliminate crossover from initial conditions to asymptotic behavior) and then using the next 2400 steps, shown in Fig. 6.  As seen in Fig. 6, $\lambda_J = 0.41$ and  $\lambda_M = 0.39$.

In addition to chaos, the renormalization-group trajectories show asymptotic strong-coupling behavior \cite{Demirtas},
\begin{equation}
\overline{J'} = b^{y_{RJ}}\, \overline{J}\,\,\text{   and  }\,\, \overline{M'} = b^{y_{RM}}\, \overline{M}\,,
\end{equation}
where the prime denotes renormalized and $y_R >0$ is the strong-coupling runaway exponent \cite{Demirtas}, as high renormaliztion-group steps.  We find here the same value of $y_R = 0.24$ for both spin glasses, which appears to be common to a large number of otherwise different spin glasses, reflecting that spin-glass order is very unsaturated order.\cite{Artun} The runaway exponent $y_R >0$ determines the compactness of the ordering by the closeness (from above) of $y_R$ to $d-1$.  In the latter case, the interface between compactly ordered regions is flat.

\section{Conclusion}

We solved, by renormalization-group theory, the Ashkin-Teller spin-glass in $d=3.$  There are 7 different phases in the calculated global phase diagram, including two chaotic spin-glass phases.  An entropic ferromagentic phase occurs, in temperature, above a spin-glass phase, which is understood by entropy consideration.  Phase reentrance and unusual reverse phase reentrance is seen.  A quasi disorder line is \textit{a priori} identified and indeed no phase transiton occurs on this line.  Thus, a global phase diagram obtains, rich in the variety of phenomena.

\begin{acknowledgments} Support by the Academy of Sciences of Turkey (T\"UBA) is gratefully acknowledged.  We thank E. Can Artun for useful conversations.
\end{acknowledgments}

\end{document}